\documentclass[aps,amsfonts,showpacs]{article}

\renewcommand{\Large}{\large}
\usepackage{amsmath}
\usepackage{graphicx}
\usepackage{subfigure}
\usepackage{captionsetup}
\usepackage[subnum]{cases}
\usepackage{color,graphics,epsfig}
\usepackage{cancel}
\usepackage{adjustbox}
\usepackage{textcomp}

\begin{document}
 
\def\be{\begin{equation}}
\def\ee{\end{equation}}
\def\bea{\begin{eqnarray}}
\def\eea{\end{eqnarray}}
\def\bml{\begin{mathletters}}
\def\eml{\end{mathletters}}
\def\b{\bullet}
\def\eqn#1{(~\ref{eq:#1}~)}
\def\no{\nonumber}
\def\av#1{{\langle  #1 \rangle}}
\def\m{{\rm{min}}}
\def\M{{\rm{max}}}

\begin{flushleft}
{\Large
\textbf{Exploiting the adaptation dynamics to predict the distribution of beneficial fitness effects }
}
\\
Sona John$^{1,\ast,\P}$,  Sarada Seetharaman$^{1,\P}$
\\
$^1$Theoretical Sciences Unit,
Jawaharlal Nehru Centre for Advanced Scientific Research, Jakkur P.O.,
Bangalore 560064, India
\\
$\ast$ 
sonajohn@jncasr.ac.in\\
$^{\P}$ These authors contributed equally to this work
\end{flushleft}

\section*{Abstract}
Adaptation of asexual populations is driven by beneficial mutations and therefore the
dynamics of this process, besides other factors, depend on the distribution of beneficial fitness effects. 
It is known that on uncorrelated fitness landscapes, this distribution can only
be of three types: truncated, exponential and power law. We performed extensive stochastic simulations to study the adaptation 
dynamics on rugged fitness landscapes, and identified 
two quantities that can be used to distinguish the underlying distribution of beneficial fitness effects.
The first quantity studied here is the fitness difference between 
successive mutations that spread in the population, which is found to decrease
in the case of truncated distributions, remain nearly a constant for 
exponentially decaying distributions and increase when the fitness distribution decays as a power law. 
The second quantity of interest, namely, the rate of change of fitness with time also shows quantitatively
different behaviour for different beneficial fitness distributions.
The patterns displayed by the two aforementioned quantities are found to hold for
 both low and high mutation rates. We discuss 
how these patterns can be exploited to determine the distribution of beneficial fitness 
effects in microbial experiments.

\section*{Introduction}
\begin{figure}[ht!]
\renewcommand{\figurename}{\textbf{Fig.}}
\centering
\includegraphics[width=1.0 \linewidth,angle=0]{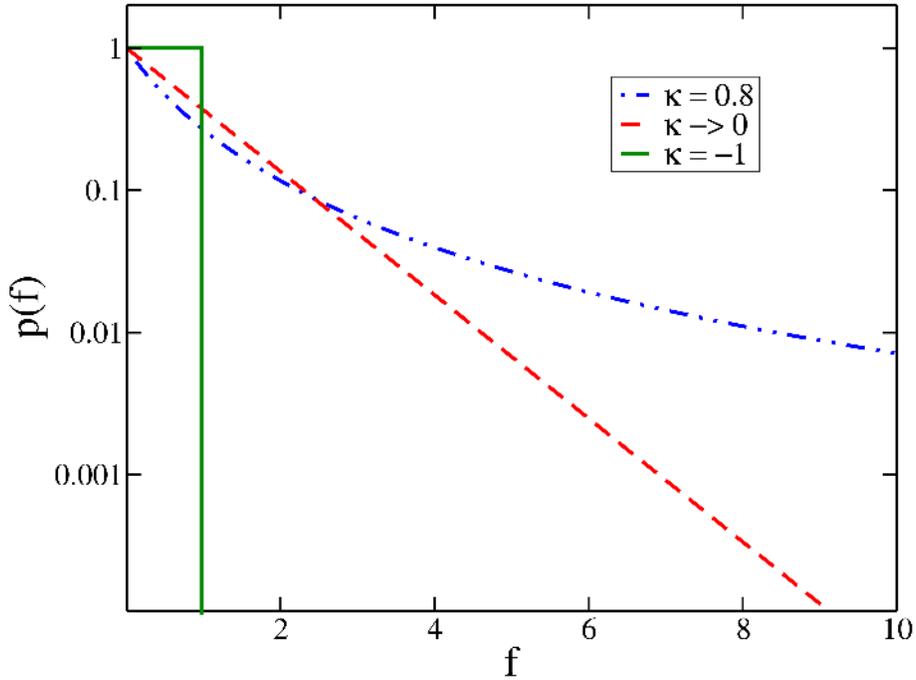}
\caption{\textbf{The figure shows the distribution of beneficial fitness effects $p(f)$ with fitness $f$
for the three EVT domains,
given by Equation Eq.~(\ref{gpd}) for various $\kappa$}. Here $\kappa$ is the tuning parameter with
$\kappa > 0$, $\kappa \rightarrow 0$ and $\kappa <0 $ 
corresponding to Fr{\'e}chet, Gumbel and Weibull domains respectively.} 
\label{fig_gpd}
\end{figure}

Microbial populations have to constantly adapt in order to survive in a changing environment. 
For example, a bacterial population exposed to a new antibiotic must evolve in order to 
exist \cite{Bull:2005b}.
In asexual populations, this process of adaptation is driven only by rare beneficial
mutations \cite{Eyrewalker:2007} which provide fitness advantage.
So in order to survive in new environment, enough beneficial mutations should be available and 
the beneficial mutations should confer sufficient fitness advantage.
While the first factor depends on the mutation rate and population size, the second factor is 
determined by the underlying fitness distributions.
Even though we have some understanding about the mutation rate of
different microbial populations, the full fitness distribution is more complex and relatively little is known about it.
But for moderately adapted populations (i.e. fitness of the wild type is high enough),
rare beneficial mutations which occur in the tail of the fitness distribution can be described by 
the extreme value theory (EVT) as proposed first by Gillespie \cite{Gillespie:1983}. The EVT
states that the extreme tail of all distributions of uncorrelated 
random variables 
(fitness, in this case) 
can be of three types only.
Depending on whether the tail of
underlying fitness 
distribution is truncated or decaying faster than a power law or as a power law, 
the EVT distribution 
would belong to 
Weibull or Gumbel or Fr{\'e}chet domain, respectively \cite{Sornette:2000}. 
All the three EVT domains can be obtained from the generalized Pareto distribution given as
\begin{equation}
p(f)=(1+\kappa f)^{-\frac{1+\kappa}{\kappa}},
\label{gpd}
 \end{equation} 
where $\kappa$ is the tuning parameter.
One example from each of the three EVT domains is shown in Fig.~\ref{fig_gpd}, which shows the 
distribution of beneficial effects $p(f)$ with fitness $f$. The three types of 
EVT domains are classified according to the value of $\kappa$.
Here negative $\kappa$ belongs to the 
Weibull domain, while $\kappa=0$ corresponds to Gumbel domain and positive $\kappa$ to 
Fr{\'e}chet domain.
Interestingly, all the three DBFEs
have been observed in experiments on microbial populations  
\cite{Sanjuan:2004a,Rokyta:2005,Kassen:2006,Rokyta:2008,Maclean:2009,Bataillon:2011,Schenk:2012,Foll:2014,Bank:2014,Rokyta:2009}. 
While the exponential distribution belonging to the Gumbel domain has been most commonly 
seen 
\cite{Sanjuan:2004a,Rokyta:2005,Kassen:2006,Maclean:2009}, in recent times, the distribution of 
beneficial mutations belonging to Weibull \cite{Bataillon:2011,Rokyta:2009} and Fr{\'e}chet \cite{Schenk:2012} 
domains have also been observed. 

Recent theoretical studies have shown analytically and numerically that qualitatively different patterns 
occur in the adaptation dynamics of populations in different EVT domains of 
DBFEs in low mutation regime
\cite{Jain:2011d,Seetharaman:2014a, Seetharaman:2014b,Seetharaman:2011}. 
Specifically, it has been shown that the fitness gain in a fixation event follows the pattern of 
diminishing returns in Weibull domain, constant returns in 
Gumbel domain and accelerating returns in Fr{\'e}chet domain, and thus indicates that 
this quantity can be used to predict the DBFE. But these observations are 
restricted to strong selection-weak mutation (SSWM) 
regime in which the genetic 
variation in the population is minimal, that is, only one beneficial mutation is present in the population 
in the time interval between its appearance and fixation \cite{Rokyta:2005}. It is then natural 
to ask whether the relationship between the adaptation dynamics and the 
DBFE mentioned above holds 
for large populations as well, where there might be more than one beneficial mutation competing for 
dominance in the population. 
The main aim of our study is to address this question and to see if the fitness gain in a
fixation event can be used for predicting the DBFE in a more general scenario.

\begin{figure}[h!]
\renewcommand{\figurename}{\textbf{Fig.}} 
\centering
\includegraphics[width=1.0 \linewidth,angle=0]{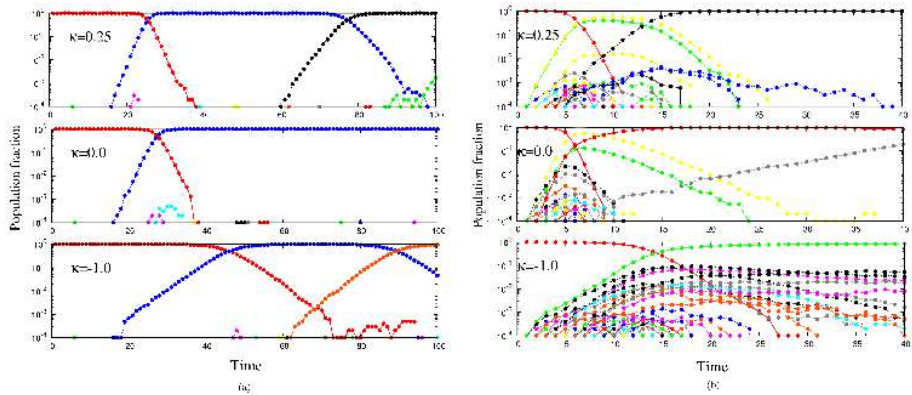}
\caption{\textbf{Population fraction of different mutant classes are shown as different 
coloured lines}. Where, (a) shows the SSWM ($N\mu=0. 1$, low mutation rate) regime
and (b) shows the high mutation ($N\mu=10$) regime
for all three EVT domains of DBFE. }
\label{sch_1}
\end{figure}

Here we are mainly concerned with the populations in which a large number of mutants are 
produced at every generation.
Hence, more than one beneficial mutation is expected to be present at the 
same time \cite{Muller:1964,Gerrish:1998,Park:2007b,Desai:2007a,Jain:2011}. 
In this case, the beneficial mutations will
compete with each other as has been observed in different experimental populations 
\cite{Visser:2006,Visser:1999,Miralles:1999,Rozen:2002}. In this high mutation
regime, as a result of the competition 
among the beneficial mutations, the rate of adaptation slows down. 
The fitness advantage due to the mutations 
that get fixed is much higher, since the availability of more mutations results in allowing only 
the best (fittest) mutation 
to get fixed \cite{Park:2010}. 
A clear comparison of the population fraction of new mutants appearing in the population for two mutation regimes 
is given in Fig.~\ref{sch_1}. 
In Fig.~\ref{sch_1}(a) 
we see that the population in the SSWM regime is more or less monomorphic with only one mutant 
present at a time in all the three EVT domains. However, in high mutation regime, population is  
polymorphic with more than one mutant 
produced in it at every generation as shown in Fig.~\ref{sch_1}(b). In fact, a large amount of 
genetic variation is observed in the case of bounded distributions corresponding to $\kappa < 0$
in Eq.~(\ref{gpd})
resulting in strong competition between the beneficial mutants.

In this work, we have used Wright-Fisher dynamics to study the adaptation 
dynamics of an asexual 
population in
high and low mutation regimes for
 the three EVT domains of DBFE. 
 The main motivation of this study is to look for quantities which can be used to distinguish between the 
 DBFEs using the
 properties of adaptation dynamics as opposed to the direct measurements of DBFEs.
 Our most important and interesting result is concerned with fitness difference between mutations that spread in 
 the population which shows qualitatively different trends in three EVT domains and thus 
 helps in distinguishing the DBFEs.

We have also studied another quantity which is the rate of change of fitness with time, 
and observed that this 
shows quantitatively different behaviour for different EVT domains of the DBFEs. Though 
some results for
the rate of change of fitness are already 
 known in the literature \cite{Park:2008}, we measured it for all the three cases 
 (Weibull, Gumbel and Fr{\'e}chet) and 
 identified that this can be used to distinguish the DBFEs in
 both SSWM and high mutation regimes.
  To obtain a 
 complete picture, a comparison of our study with the existing literature is given in Table \ref{table} below.

 \begin{table}[htbp!]
 \begin{adjustbox}{max width=\textwidth}
 \begin{tabular}{|c||c|c|} \hline
Quantities & DBFE domains: Low mutation regime & DBFE domains: High mutation regime  \\ \hline
 \begin{tabular}{c}
  \\ 
 $\overline{ \Delta{f_{step}}}$  \\ 
  $\bar{\cal F}(t)$ 
\end{tabular}
 &
\begin{tabular}{c|c|c}
 Weibull & Gumbel & Fr{\'e}chet \\ \hline
 \cite{Seetharaman:2014a} & \cite{Seetharaman:2014a} & \cite{Seetharaman:2014a} \\ \hline
 this study & \cite{Park:2008}& \cite{Park:2008} 
\end{tabular}
&
\begin{tabular}{c|c|c}
 Weibull & Gumbel & Fr{\'e}chet \\ \hline
 this study&this study&this study\\ \hline
 this study&\cite{Park:2008}&this study
\end{tabular} \\ \hline
\end{tabular}
\end{adjustbox}
\caption{Here, $\overline{ \Delta{f_{step}}}$ is the average fitness difference between the present
leader and the new beneficial mutation that gets established and $\bar{\cal F}(t)$
is the rate of change of fitness.}
\label{table}
\end{table}

 We also measured quantities like the genetic variation and the number of 
 mutations in the most populated sequence. All of these quantities 
 are discussed in Results section.
 We suggest that the distinct trends shown by the above mentioned quantities can be used to predict DBFEs 
from experimental studies on adaptation.
 The relevance of our 
 work to 
 experiments is also explored in Discussion section.

 \section*{Materials and Methods}
 
 We track the dynamics of a population of self-replicating (asexual), infinitely long binary sequences of 
 fixed size using the standard Wright-Fisher process \cite{Park:2007b,Park:2010}. 
In our work, the population size is held constant at $N=10^4$, unless specified otherwise and the total
mutation 
probability (beneficial and deleterious) per sequence is given by $\mu$. 
Every occupied sequence is counted as a {\it class} and labelled when it arises in the population. 
Initially, the whole population is in class $1$ whose fitness is fixed and specified in every simulation run.
We have used the term leader to refer to the class whose normalised probability of reproduction
(product of population fraction and fitness) is greater than half. 
In that case, clearly, class $1$ is the initial leader since the whole population is localized there. 
At every time step, out of $N$ sequences, $m_t$ are chosen from a binomial distribution with mean 
$N\mu$ as mutants. Every mutant produced increases the number of classes in the population by one, and 
with time, the mutants may produce their own set of further mutants. 
 The population fraction of each class may grow or go extinct, as can be observed in Fig.~\ref{sch_1}. At
 any time $t$, the number of classes present in the population is given by ${\cal N}_{c}{(t)}$, and the population 
 size and fitness of each class, $i$, where $1\le i\le{\cal N}_{c}$, is denoted by $n(i,t)$ and $f(i)$, 
 respectively. The normalized probability of each class at every time step, $\tilde{p}(i,t)$ 
 contributing offspring to 
 the population at the next time step, depends on the population size of the class at the present 
 time step and the fitness of the class as
\begin{equation}
{\tilde{p}(i,t)}=\frac{{n(i,t)}~ f(i)}{\Sigma_{j=1}^{{\cal N}_{c}(t)}{n(j,t)} f(j)}. 
 \label{pro_i}
\end{equation}
Note that though the fitness of the class is the same as long as it persists in the population, its size may vary at 
every time step, thus changing its probability of reproduction as given by Eq.~(\ref{pro_i}). 
The different classes are populated in the next time step based on the multinomial distribution
\begin{equation}
P(n(1,t'),n(2,t'). . n({\cal N}_c,t'))= N! \prod_{j=1}^{{\cal N}_c(t)} 
 \frac{[\tilde{p}(j,t)]^{n(j,t)}}{n(j,t)!}
 \label{multi}
\end{equation}
where $t'=t+1$. The above equation is subject to the constraint $\Sigma_{j=1}^{{\cal N}_c(t)} {n(j,t')}=N$. 
In our simulations, we implement Eq.~(\ref{multi}) along with the above constraint by converting Eq.~(\ref{multi}) 
to a binomial distribution for every class, $1 \le i < {\cal N}_c(t)$ as 
\begin{equation}
n(i,t') = {\tilde N(i) \choose n(i,t)} q(i,t)^{n(i,t)} (1-q(i,t))^{\tilde N(i)-n(i,t)}
\label{bin}
\end{equation} 
We set the population size of the last class as $n({\cal N}_c(t),t')=N-\sum_{i=1}^{{\cal N}_c(t)-1}~n(i,t')$.
In Eq.~(\ref{bin}),
\begin{equation}
q(i,t)=\frac{\tilde{p}(i,t)}{ \Sigma_{j=i}^{{\cal N}_c(t)} \tilde{p}(j,t)}
\end{equation}
and $\tilde N(i)=N-\Sigma_{j=1}^{i-1}n(j,t)$. 

At every time step, once the classes are populated based on the algorithm described above, 
$m_t$ sequences are chosen as
mutants based on the binomial distribution with mean $N\mu$. Every new mutant class that appears in the population reduces 
the population size of the class in which it arose by one. 
In our work, we have varied $\mu$ to access both the SSWM (low mutation) and 
the high mutation regime. 
In our simulations unless specified otherwise, $N\mu=0. 01$ in low (SSWM) and $N\mu=50$ in high mutation regimes.

A new class is assigned to  each mutant and its fitness is chosen from a generalized Pareto 
distribution \cite{Sornette:2000} given in Eq.~(\ref{gpd}). 
 The advantage of using Eq.~(\ref{gpd}) is that we can access all three EVT domains of DBFE by changing $\kappa$. The distributions 
 whose $\kappa<0$ belong to the Weibull domain, while $\kappa=0$ belong to the Gumbel domain, and $\kappa>0$ 
 belong to the Fr{\'e}chet domain, respectively. The frequency distribution of beneficial effects 
 $p(f)$ for various values 
 of $\kappa$ is shown in Fig.~\ref{fig_gpd}. 
 The upper bound $u$ for the distributions chosen from Eq.~(\ref{gpd}) is 
 infinity when $\kappa\ge0$ and equals $-1/\kappa$ for $\kappa<0$. 
 In this work, the fitness of the mutants is independently chosen from Eq.~(\ref{gpd}) thus making the fitness of the mutant, 
 $F_m$ an uncorrelated variable, which may be greater or smaller than the parent fitness, $F_p$. 
 We have analyzed the 
 results to see how they vary between the three EVT domains and different mutation rates.

 In the allocation of the 
 fitness to any mutant, our work differs from 
 the other works on clonal interference \cite{Park:2007b, Park:2010} wherein the fitness of the mutant is hiked above 
 the parent fitness by the selection coefficients ($s$) which may be 
 held constant or chosen from a distribution as $F_m=(1+s)F_p$. 
Unlike the model we have used in this work (as explained above), in this case there is a 
 strong correlation between the mutant fitness $F_m$ and the parent fitness $F_p$. 
 In those 
 cases, the mutant fitness is always greater than the parent fitness and on an average, a double or higher mutant is fitter 
 than a single mutant. 
 This is in contrast with our work since in ours, as the fitness of the parent increases, the number of better mutants available 
 decreases thus producing different patterns for the fitness increment in each EVT domain.

In our model, whenever a mutant class goes extinct,
the classes below
it is moved up, and the number of classes in the population is reduced by one. 
  The normalized probability of reproduction given in Eq.~(\ref{pro_i}) of a 
  class exceeding half corresponds to a leader change. The new leader 
  determined now belongs to the class whose normalized probability exceeded half.
  We have also explored other criteria for defining the leader as the 
  most populated class and find that our main results are robust with respect to the change in 
  criteria (data not shown).
  
  Every change of leader is counted as a 
  \textit{step}. In the high mutation regime the population is spread 
  over many sequences and a sequence can 
  produce two or more mutants each of which  may become leaders at different time steps. 
  However, in the SSWM 
  regime, the whole population is localized at a single sequence with a fixed fitness and can only move to a different
  sequence with higher fitness one mutation away. Thus every new leader arises from the previous leader, 
  as can be observed in Fig.~\ref{sch_1}(a). When a better sequence appearing in the
  population does not get lost due to 
  genetic drift, it quickly gets fixed. Further mutations that may lead to future leaders appear in this 
  genetic background. 
  The change in the 
  fitness of the population is the same as the change in fitness of the leader. In this case, every move of the population (leader) 
  from one sequence to another is termed a step in the adaptive walk 
 \cite{Wilke:1999,Orr:2003a,Rosenberg:2005,Kryazhimskiy:2009}, whereas in high mutation regime, 
 the population is polymorphic as can be seen from Fig.~\ref{sch_1}(b) and the leader change is not obvious from the figure. 
  
  Various quantities like the fitness difference between successive leaders and the average number of mutations in the leader 
  are averaged only over the walks that take the step. Other quantities like the number of classes present at any time point 
  and the rate of change of fitness are averaged over all time steps in that simulation run.  
  
   In this paper, the total number of iterations is $10^5$ in every simulation run and the dynamics is tracked 
   for finite time limit of $10^4$ generations which we shall refer to as $t_{max}$. 
   In this time span, the maximum fitness value, $f_{max}$ that arises in 
   the population can be calculated as 
  \begin{equation}
  t_{max} N\mu \int_{f_{max}}^u p(f) df=1
   \label{f_max}
  \end{equation}
  where $u$ is the upper limit of the fitness distribution equalling  (-1/$\kappa$) 
for bounded distributions and infinity for unbounded ones \cite{Sornette:2000}. From the above integral, we get
\begin{equation}
 {f_{max}} =\frac{ (t_{max}N\mu) ^\kappa -1}{\kappa}. 
 \label{fmax}
\end{equation}

\section*{Results}
\subsection*{The number of classes in the population}
\begin{figure}[h!]
\renewcommand{\figurename}{\textbf{Fig.}} 
\centering
\includegraphics[width=1.0\linewidth,angle=0]{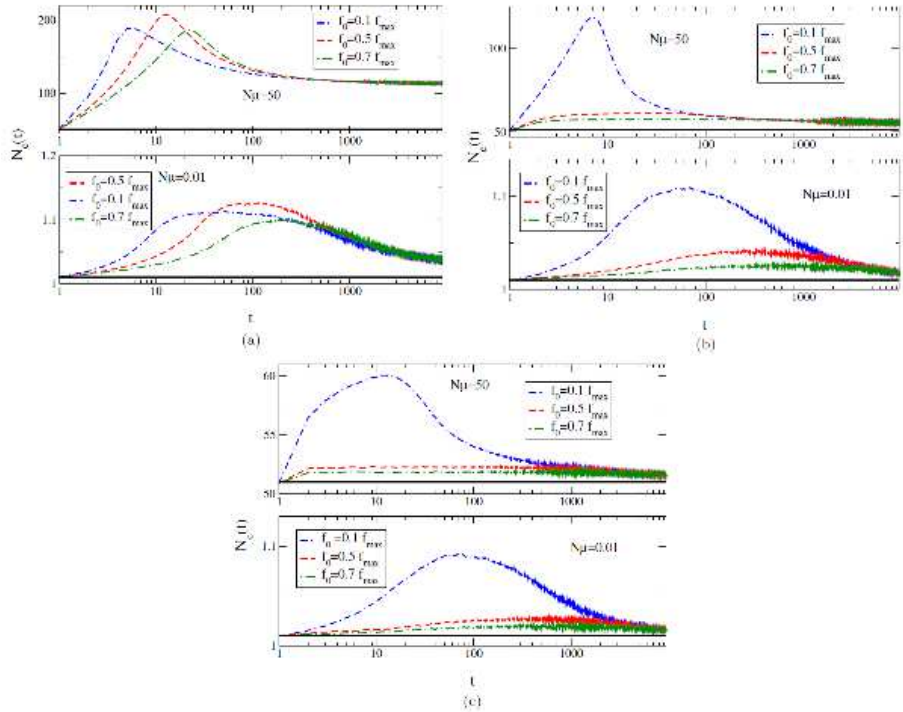}
\caption{\textbf{The plot shows the average number of classes in the population as function of time for various initial 
fitnesses. } The fitnesses are chosen from Eq.~(\ref{gpd}) with (a) $\kappa=-1$ (b) $\kappa\rightarrow0$ and (c) $\kappa=1/4$. 
For each $\kappa$ value, the plot shows ${\cal N}_c{(t)}$ in both the high mutation 
   (top panels) and low mutation  
 (bottom panels) regimes. The straight line in all plots shows $N\mu+1$. }
\label{Classes}
\end{figure}

For a population of fixed size, the number of classes in the population is expected to increase with the mutation rate. 
The average genetic variation defined here as the average number of 
classes (${\cal N}_c$) present in the population is shown in Fig.~\ref{Classes} 
for all the three domains of DBFE. 
The top and bottom panels of the figure show the data corresponding to the high and low 
 mutation regimes respectively. 
In both the mutation regimes, we see that the average number of classes increases during the initial 
 time steps and decreases at later times when the classes with lower fitness are eliminated 
 by the fitter ones.
  The maximum number of classes existing 
in the population 
for the first case, as shown in Fig.~\ref{Classes}(a), does not belong to the lowest initial fitness, but to a slightly higher 
initial fitness. This could be because when 
the initial fitness is low, its class is quickly replaced by a fitter mutant and all further mutants arise on 
this new background must compete with this fitter class. 

In the low mutation regime, the population for the most time is localized at a single sequence and produces $N\mu$ 
mutants at every time step. So in this case, the average number of classes approach a constant $N\mu+1$ at large times 
as can be seen in the bottom panels of  Fig.~\ref{Classes}. These panels also indicate that the value of this constant 
increases with decreasing $\kappa$. This is because in the case of bounded distributions with 
$\kappa < 0$, the fitness of beneficial mutant produced is expected to be closer to the parent 
fitness. In other words, mutations are nearly neutral and thus it takes longer time to take over the population 
as shown in 
Fig.~\ref{sch_1}(a). This results in a larger number 
of mutants in Weibull domain which can be observed in the bottom panel of Fig.~\ref{Classes}(a). 
We can clearly see from the top panels of Fig.~\ref{Classes}
that number of classes increases with decreasing $\kappa$ even in high mutation regime. Also, the average number 
of classes present at a time is much higher in this regime. 
This makes sense because the fitness of the classes belonging to $\kappa=-1$ cannot be very different from each other (can only 
vary between $0$ and $1$) which makes it possible for many of them to exist in the population. The 
maximum fitness of the 
classes belonging to $\kappa=1/4$ distribution will, on an average be much higher than all others (since the distribution is 
unbounded with a fat tail), thus out-competing the others in the population. 

\subsection*{Number of mutations in the leader}

\begin{figure}[ht!]
\renewcommand{\figurename}{\textbf{Fig.}} 
\centering
\includegraphics[width=1.0 \linewidth,angle=0]{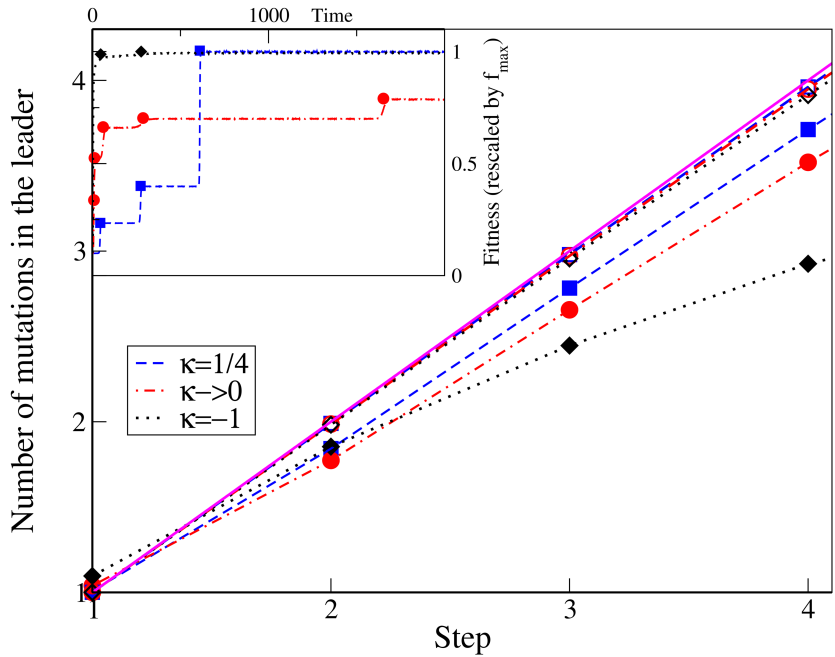}
\caption{\textbf{The main plot shows the number of mutations in the leader at any step for various $\kappa$ and mutation rates. }
The simulation data are represented by points while the broken lines connect the data points. 
The solid line shows $y=x$. In the inset, from a single simulation run, the fitness of the whole population as 
a function of time is shown by broken lines and the fitness of the leader whenever the leader changes is shown by symbols. 
}   
\label{m_no}
\end{figure}

 In the low mutation regime, the average number of mutations in the leader is expected to be
 very close to the step number since the genetic variation in the population is low and any mutation that escapes drift 
 quickly takes over the population \cite{Gillespie:1983}. We verify this point via simulations as depicted in Fig.~\ref{m_no}. 
 We find that the mutation number equals the step in all the three EVT domains of the DBFE in the low mutation 
 regime for the initial 
  steps. 
However in the high mutation regime, the number of mutations in the leader of any  step differs between the three 
DBFE domains. When the mutation rate is increased, the genetic 
 variation of the population and the significance of clonal interference also increases. In the 
 high mutation regime, 
 the number of mutations in the leader is found to be less than the step number in all the three DBFE 
 domains. This is because there is a 
 chance that different mutants originating from the same parent class can become the leader of the population 
 at different times. 
 This decrease from the step number is the minimum for the fat-tailed distributions and maximum for the 
 truncated ones, as 
shown in Fig.~\ref{m_no}. This result is consistent with the number of classes present in the population as discussed in the 
previous section. 
 In the Fr{\'e}chet domain, since the clonal interference is minimal, mostly a mutant originating from the 
present leader will become the next one. In the Weibull domain, due to the large number of classes
present in the 
population, mutants originating from the same class can become the leaders at different time points.          
 
\subsection*{Fitness and fitness difference}

From our simulations, we find that the average fitness of the first mutant fixed in the population, 
$\bar{f_1}$ increases linearly with the initial fitness, $f_0$ for all $\kappa$ in the low mutation regime and for $\kappa\ne 0$ 
in the high mutation regime. So we can write
 \begin{equation}
 {\bar{f_1}}=a_\kappa^{(N\mu)} f_0+b_\kappa^{(N\mu)}
 \label{f1}
\end{equation}
where the coefficients $a_\kappa^{(N\mu)}$ and $b_\kappa^{(N\mu)}$ are constants. In the low mutation regime,
where the population for most times is monomorphic, the adaptive walk model has been used to 
 analytically obtain 
 the fitness at the first step, $ \bar{f_1}$ as 
\cite{Jain:2011d, Seetharaman:2014a} 
\begin{equation}
 \bar{f_1}=\int_{f_0}^u df~  T(f \leftarrow f_0) f
 \label{theo_f1}
 \end{equation}
  where the transition probability
 \begin{equation}
   T(f \leftarrow f_0)  =  \frac{(1-e^{-\frac{2 (f-f_0)}{h}})p(f)}{\int_{f_0}^u dg~ \left(1-e^{-\frac{2 (g-f_0)}{f_0} }\right )p(g)} . 
\label{Tfull}
  \end{equation}
 In this model, from Eq.~(\ref{theo_f1}), the coefficient $a_\kappa^{(N\mu\ll1)}$ was obtained as $0. 33, 1. 0 \text{ and }
 1. 6$ for $\kappa=-1,0, \text{ and } 1/4$ respectively. The corresponding $b_\kappa^{(N\mu\ll1)}$ for the aforementioned 
 $\kappa$ were $0. 66, ~ 2. 0  \text{ and } 1. 89$  \cite{Seetharaman:2014a}. In the high mutation regime 
 where the adaptive walk model is not applicable,  we obtained the values for the coefficients in Eq.~(\ref{f1}) numerically. 
 We find that for large $f_0$, $a_\kappa^{(50)}$ equals $ 0. 004
 \text{ and } 1. 5 $ and $b_\kappa^{(50)}$ equals  $0. 99
 \text{ and } 9. 1 $ for $\kappa=-1 \text{ and } 1/4$ respectively.

\begin{figure}[h!]
\renewcommand{\figurename}{\textbf{Fig.}} 
\includegraphics[width=1.0 \linewidth,angle=0]{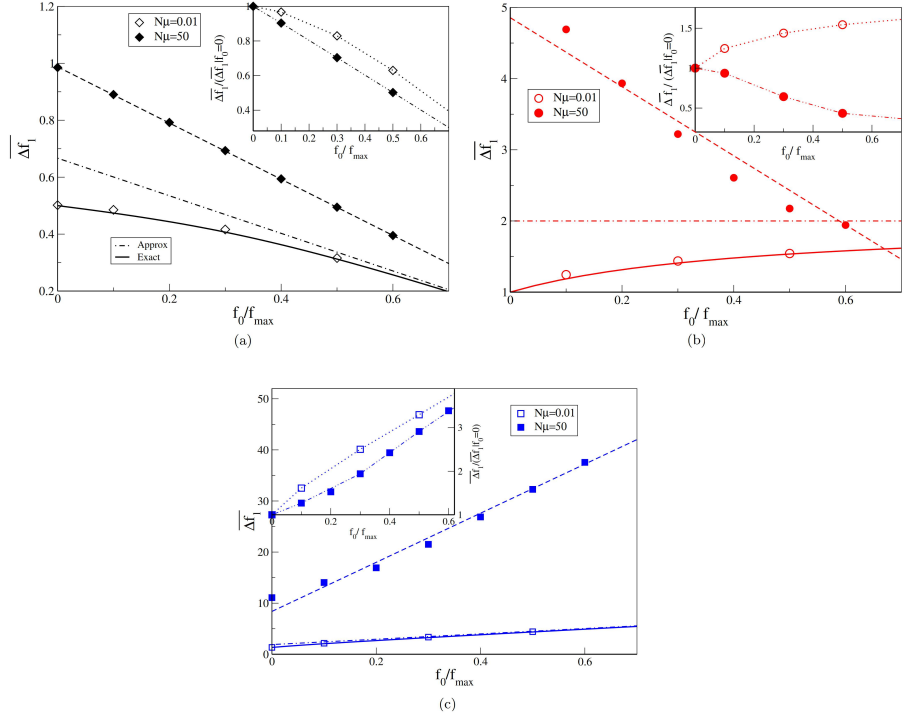}
\caption{\textbf{The main plot shows the fitness difference at the first step as a function of the initial fitness for 
various $N\mu$}. The fitnesses are chosen from Eq.~(\ref{gpd}) with (a) $\kappa=-1$ (b) $\kappa\rightarrow0$ and (c) $\kappa=1/4$. 
The solid lines in the main plot are obtained by numerically evaluating the integral given by Eq.~(\ref{theo_f1}), while the dotted 
lines are the approximate results that can be obtained  
for the results when the initial fitness is high, in the low mutation regime. The broken lines for $\kappa\ne0$ are lines 
of best fit as mentioned in the text. The broken line for $\kappa\rightarrow0$ is 
used for connecting the data points. 
 The inset shows the fitness difference at the first step as a comparative measure of the fitness difference obtained at 
 the first step when $f_0=0$. Here, the lines are  used for connecting 
 the data points. }
\label{f1-f0_allk}
\end{figure}

 The interesting result from our work is that, irrespective of the number of mutants produced in the population, 
 the difference $\overline{\Delta{f_{step}}}=\bar{f_1}-f_0$  between the fitness of the first step and 
 the initial fitness
 displays different qualitative trends: increases for positive $\kappa$, approaches a constant 
 when $\kappa=0$ and decreases for negative $\kappa$, as shown in Fig.~\ref{f1-f0_allk} 
 and \ref{rf0}. 
 
We can better understand these increasing and 
decreasing trends 
by the following heuristic argument. In both the low and high mutation regimes, for large $f_0$, the fitness 
at the first step $f_1$ increases linearly with the initial fitness as given in Eq.~(\ref{f1}) and so, we can write the selection 
coefficient defined as 
 the relative fitness difference, at the first step as 
 \begin{equation}
s=\frac{\bar{f_1}-f_0}{f_0}=\frac{(a_\kappa^{(N\mu)}-1)f_0}{f_0}+\frac{b_\kappa^{(N\mu)}}{f_0},~~~\text{for all} ~~~~
\kappa \text{ , } N\mu
  \label{s}
 \end{equation}
In an adapting population, since the fitness of the first step is greater than the initial fitness, 
the selection coefficient is always positive. 
As the fitness distributions 
belonging to the Fr{\'e}chet domain are unbounded with fat tails, high $f_0$ values can be considered in which case, the second 
term on the right hand side (RHS) of Eq.~(\ref{s}) can be ignored and we can write $s \approx
(a_\kappa^{(N\mu)}-1) > 0 $. Thus for $\kappa>0$, $a_\kappa^{(N\mu)}>1$ and therefore it follows 
that the fitness difference at the first
step increases with $f_0$. 
On the other hand, since the distribution belonging to the Weibull domain are truncated, we can invoke the following 
inequality to explain the decrease in fitness difference with increasing $f_0$:
\begin{equation}
 \bar{f_1}-f_0<u-f_0,
 \label{u}
\end{equation}
where $u$ is the upper limit of the fitness distribution. With increasing $f_0$, the RHS of the above 
equation decreases which shows that as the initial fitness increases, $\bar{f_1}-f_0$ has to necessarily decrease. 
Thus the qualitative trends discussed above appear to be determined 
by the behaviour of the tail (bounded/unbounded), and not by the details of the model.

  Also, it is interesting to note that while the data points for the exponentially decaying distribution 
  ($\kappa=0$) increase 
and seem to be approaching a constant in the low mutation regime, the data in the high mutation 
regime seems to be reducing to approach the same constant. Our simulation results  shown in Fig.~\ref{f1-f0_allk}
not only match the predicted theoretical values and  validate the claim of  different qualitative trends in each 
EVT domain in the SSWM regime but also show that the trends hold 
irrespective of the number of 
mutants produced in the population. This result suggests that the
qualitatively different trends of the fitness difference 
(increasing, constant and decreasing with initial fitness in the Fr{\'e}chet, Gumbel and Weibull domain, respectively) can be  
used to distinguish between the EVT domains in a more general scenario. 

Though the fitness difference at the first step is greater in the high mutation regime,
when compared with the results in the low mutation regime, when we look at the fitness difference at the 
first step scaled by the fitness difference obtained when the initial fitness is zero (insets of 
Fig.~\ref{f1-f0_allk}), we see that this increase is slower in the high mutation regime compared to the results 
obtained in the low mutation regime. This indicates that as the mutation rate increases, though the number of 
mutants accessed is higher, the difference in fitness compared to a lower initial fitness is not proportionally 
higher and is in fact lower for all the fitness distributions. 

\subsection*{Rate of change of fitness with time}

\begin{figure}[ht]
\renewcommand{\figurename}{\textbf{Fig.}} 
\centering
\includegraphics[width=1.0 \linewidth,angle=0]{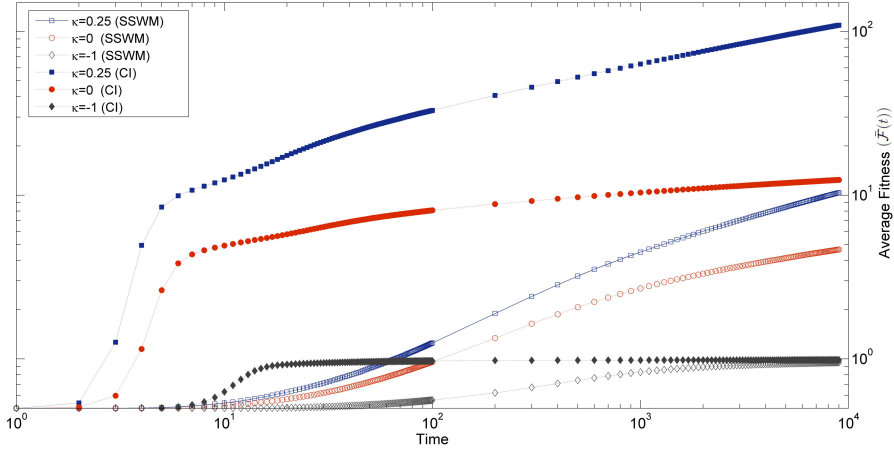}
\caption{\textbf{Figure shows the average fitness increase with time for three different 
values of $\kappa$ in the SSWM regime ($N\mu=0. 01$), and in high mutation
regime($N\mu=50$). } 
In all the cases, the population starts with the same initial fitness $f_0=0. 5$.}   
\label{fit_fig6}
\end{figure}

Besides the fitness increment at a fixed event of leader change, we also measured the fitness 
as a function of time as shown in Fig.~\ref{fit_fig6}. We observed that even
though the fitness increases with time in all the three EVT domains, the rate at which 
the fitness increases depends strongly on the DBFE. 
This rate has an initial fast transient phase, after which it slows down. 

The initial transient phase is strongly dependent on the initial condition as well as the mutation rate as 
shown in \ref{RA_ic}. The increase in fitness is fastest for the lowest initial condition, 
but it approaches the same fitness value as in the 
case of higher initial fitness in few generations. The time taken for populations of different initial fitness to 
reach the same fitness value depends on the mutation rate: for $N \mu \gg 1$, it takes about $20$ generations, 
whereas for  $N \mu \ll 1$,  it is approximately $200$ generations. Even after this transient phase, 
the rate of increase in average 
fitness ($\bar{\cal F}(t)$) with time depends on the mutation rate as shown in Fig.~\ref{fit_fig6}. 
This is because of the fact that, when a large number of mutations are available at the same time, 
a highly fit mutant can invade 
the population and give a large fitness increment. So the fitness of a highly fit 
mutant sequence would be greater in 
the high mutation regime compared to the one in low mutation regime. The
maximum fitness value reached in $9000$ generations, in the case of Fr{\'e}chet distribution is about $10$ times more
for high mutation regime, which is consistent with the 
expectation from Eq.~(\ref{fmax}). 
Even beyond this point we noticed that the fitness is still 
increasing. In the same way, 
Gumbel distribution also shows a significant increase in 
maximum fitness reached in high mutation regime compared to the SSWM regime 
(about $4$ times). 
Here also we found that the fitness is still increasing beyond the time point till which we tracked 
the dynamics.
The bounded distribution (Weibull) reaches 
near the upper bound in SSWM and evolves slowly. But the fitness reaches a fitness plateau in high mutation regime and 
rate of adaptation becomes zero as can be seen in Fig.~\ref{fit_fig6}.

\begin{figure}[ht]
\renewcommand{\figurename}{\textbf{Fig.}} 
\centering
\includegraphics[width=1.0 \linewidth,angle=0]{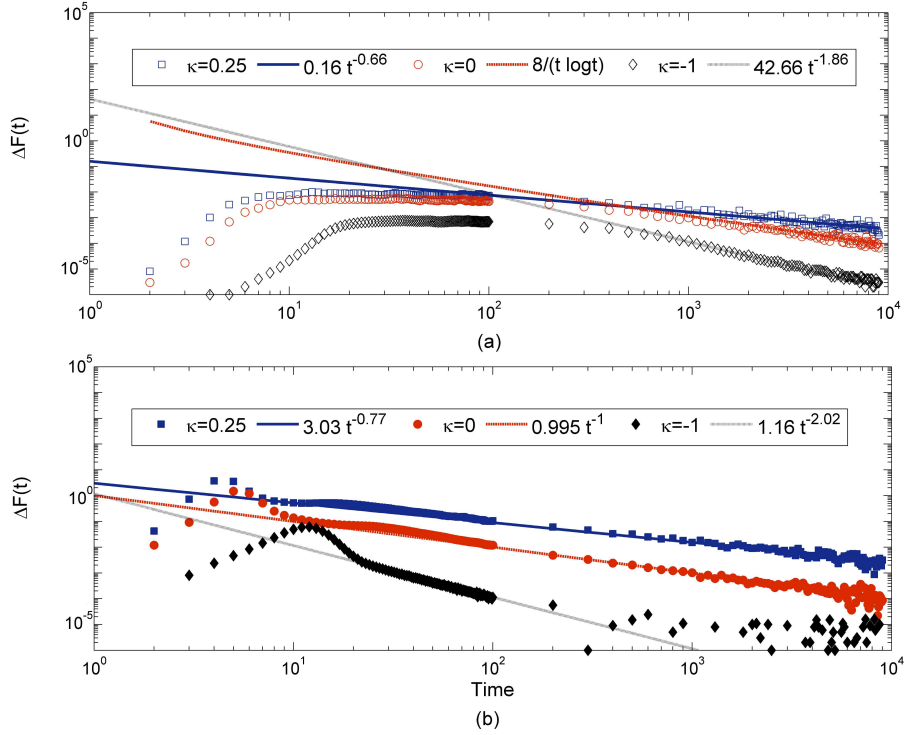}
\caption{\textbf{Figure shows the fitness increment in each time step for three different 
values of $\kappa$ in two mutation regimes (SSWM and high mutation). } In each 
case the data is fitted
with the theoretically expected function given in Eq.~(\ref{alpha}), except for exponential distribution for which 
we used the theoretical prediction by Park and Krug \cite{Park:2008}. In all the cases, the population 
starts with the same initial fitness $f_0=0. 5$. }    
\label{RA_fig2}
\end{figure}

From this we observe that the rate of change of fitness depends strongly on the properties of the underlying 
DBFE, which suggests that looking at this quantity can help us in distinguishing the DBFEs. 
So we measured the fitness increment defined as 
\begin{equation}
\Delta \bar{\cal {F}}(t) = \langle \bar{\cal{F}}(t+1) - \bar{\cal{F}}(t) \rangle
\label{RA}
\end{equation}
at each step. 
The $\Delta \bar{\cal {F}}(t)$ initially increases, then slowly decreases and settles down 
to a zero as shown in Fig.~\ref{RA_fig2}. If we denote this function as 
\begin{equation}
 \Delta \bar{\cal {F}}(t) = \frac{A}{t^{\alpha}}
 \label{alpha}
\end{equation}
 where $A$ is a constant and the exponent $\alpha$ can be used to distinguish the 
 DBFE, since, as explained below, exponent $\alpha$ is found to be greater (smaller) than one in 
 Weibull (Fr{\'e}chet) domain, but is close to one in Gumbel domain.

In the SSWM regime, 
 from Fig.~\ref{RA_fig2}(a), we can see that each type of DBFE considered shows a different rate of decay. Weibull
 domain has a faster decay with
 $\alpha=1.86$, Gumbel domain has $\alpha \approx 1$ \cite{Park:2008} and Fr{\'e}chet domain $\alpha=0.66$ 
 \cite{Park:2008}. We observed that the same trend holds in high mutation rate regime as well, where
 $\alpha$ values are slightly larger in all cases. In this regime also $\alpha=2.02$, $1$ and $0.76$ 
 for Weibull, Gumbel and Fr{\'e}chet domains respectively
 as shown in Fig.~\ref{RA_fig2}(b). In the high mutation regime, in the case of Weibull distributions fitness reaches a plateau
in few generations, after which its rate of change goes to zero, as can be observed in 
Fig.~\ref{RA_fig2}(b). 
The theoretical prediction
 for the fitness at every time step for the unbounded distributions belonging to the Gumbel and Fr{\`e}chet domains was 
 obtained by Park and Krug \cite{Park:2008} in the low mutation regime.
 The comparison of our simulation data
 with these predictions shows a very good agreement in Gumbel domain and in Fr{\'e}chet domain (up to a constant). 
In this work, 
we have considered the bounded distribution also and observed that its rate of decrease is faster with an 
exponent greater than one, 
which was not considered in the previous studies. 
We observed that even in high mutation regime, the 
exponent $\alpha$ shows the same behaviour. 
In this regime the rate of 
change of fitness has been calculated only for exponential distribution belonging to the
Gumbel domain \cite{Park:2008} and their prediction matches with our data. 
In this work, we have obtained a complete picture by studying the rate of change of fitness 
numerically for the other two EVT domains as well.

Thus, the second main finding from our study is that in all DBFEs, the fitness 
difference at each time step
decreases with time as given by Eq.~(\ref{alpha}) and we can distinguish between 
the three EVT domains of DBFEs by looking at the exponent $\alpha$.
A comparison of our results with the existing literature
is given in Table \ref{table}.

\section*{Discussion}

The main purpose of our work is to determine the quantities which can be used to 
distinguish the 
different extreme value domains of the DBFE. 
Previous studies \cite{Seetharaman:2014a,Seetharaman:2011} 
have found that in an adapting population, the fitness gain at each fixation event shows qualitatively 
different trends in the three 
DBFE domain, when the number of mutants produced in the population is much less than one at every generation 
($N\mu\ll1$). 
The focus of this work is to explore the parameter regime in which the number of mutants produced is much above one 
($N\mu\gg1$). 
When the mutation rate is high, the population becomes polymorphic and the better mutants existing in the population compete 
with each other. From our study we have observed that the qualitative trends found for fitness difference when
a new mutation establishes
in the low mutation regime hold 
irrespective of the number of mutants produced. Thus this study suggests that fitness difference between the successive 
mutations that spreads in the 
population is a very important and robust quantity that can be used to predict the DBFEs in a more general scenario.

From our simulations, we see that 
 as the initial fitness is increased the 
fitness difference at the first step given by  $\overline{\Delta f_{step}}$ reduces, approaches a constant or
increases with initial fitness in the Weibull, Gumbel and Fr{\'e}chet domains, respectively.
We can understand these trends 
by a heuristic reasoning as discussed in detail in the Results section.
This argument explains the increase in  $\overline{\Delta f_{step}}$ with $f_0$ for unbounded power law distribution
and shows that the trends are determined 
by the behaviour of the tail (bounded/unbounded), and not by the details of the model.

Another important measure in understanding the dynamics of adaptation is the rate at which it occurs. 
Most of the previous studies which measured the adaptation rate have only 
considered exponentially distributed fitness distributions \cite{Gerrish:1998,Park:2007b,Desai:2007a,Park:2010,Campos:2010}. 
A previous study by Park and Krug \cite{Park:2008} also considered DBFEs belonging to Fr{\'e}chet
domain but only in the SSWM regime (see Table \ref{table}).
In this work, we have extended the previous studies by numerically measuring the rate of 
change of fitness for bounded distributions also.
We have measured the rate of change of fitness in all the three EVT domains
of the DBFE in both low and high mutation regimes. 
We observed that in all the cases, the rate of change of fitness decreases with time as 
$\sim t^{-\alpha}$, 
where $\alpha > 1$ for Weibull, 
$\alpha \approx 1$ for Gumbel \cite{Park:2008} and $\alpha < 1$ for Fr{\'e}chet domains \cite{Park:2008}.

Experimentally, the distribution of beneficial fitness effects can be inferred by two methods. In the first method, 
mutations are introduced in the wild type sequence and those that confer a fitness advantage are separated and their distribution 
of fitness effects are determined. By this method, DBFE belonging to all the EVT domains have been observed \cite{Sanjuan:2004a,
Rokyta:2005,Kassen:2006,Rokyta:2008,Maclean:2009,Bataillon:2011,Schenk:2012,Foll:2014,
Bank:2014,Rokyta:2009}. In contrast, here we focus on learning about DBFE via
adaptation dynamics. Though many works have tracked the dynamics of 
the population during adaptation \cite{Rokyta:2005,Schoustra:2009,Maclean:2010,Gifford:2011,Sousa:2012}, in most of them only 
the selection coefficient of the mutant fixed was measured. 
In our study, we have observed that the selection 
coefficient as given by Eq.~(\ref{s}) always decreases, with the increasing initial fitness or increasing steps as 
shown in \ref{fig5}. Hence this quantity is not useful to distinguish between the EVT domains. However, 
from our study we observe that the fitness 
difference between steps shows different patterns depending on the EVT domain of the DBFEs in both the high and low mutation 
regimes and can be used to distinguish between the EVT domains.

In this work, we have numerically shown that the fitness returns in 
each EVT domain is very 
robust and holds even when the number of mutations produced is large ($N\mu\gg1$). 
Fitness difference can be measured in 
experiments, for example as in \cite{Maclean:2009}.
We suggest that experiments can predict the EVT domain of DBFE by measuring the fitness difference between 
successive mutations fixed in the population, or even from the fitness of the first 
mutation, when the initial 
fitness is varied. However currently experimental studies that measure both fitness and DBFE in the 
same study are not available but it is highly desirable to have such studies to test our predictions.

\section*{Acknowledgments}

We thank K. Jain for many useful discussions that helped us in this work and suggesting the heuristic 
argument in the discussion.
We also thank J. Krug for bringing 
references \cite{Foll:2014,Bank:2014} to our attention. We thank V. Yalasi for helping us 
to improve the figure quality.

\newpage

\section*{Supplementary Figure}
\setcounter{figure}{0}

\begin{figure}[ht!]
\renewcommand{\thefigure}{\textbf{S1 Fig.}}
\includegraphics[width=1.0 \linewidth,angle=0]{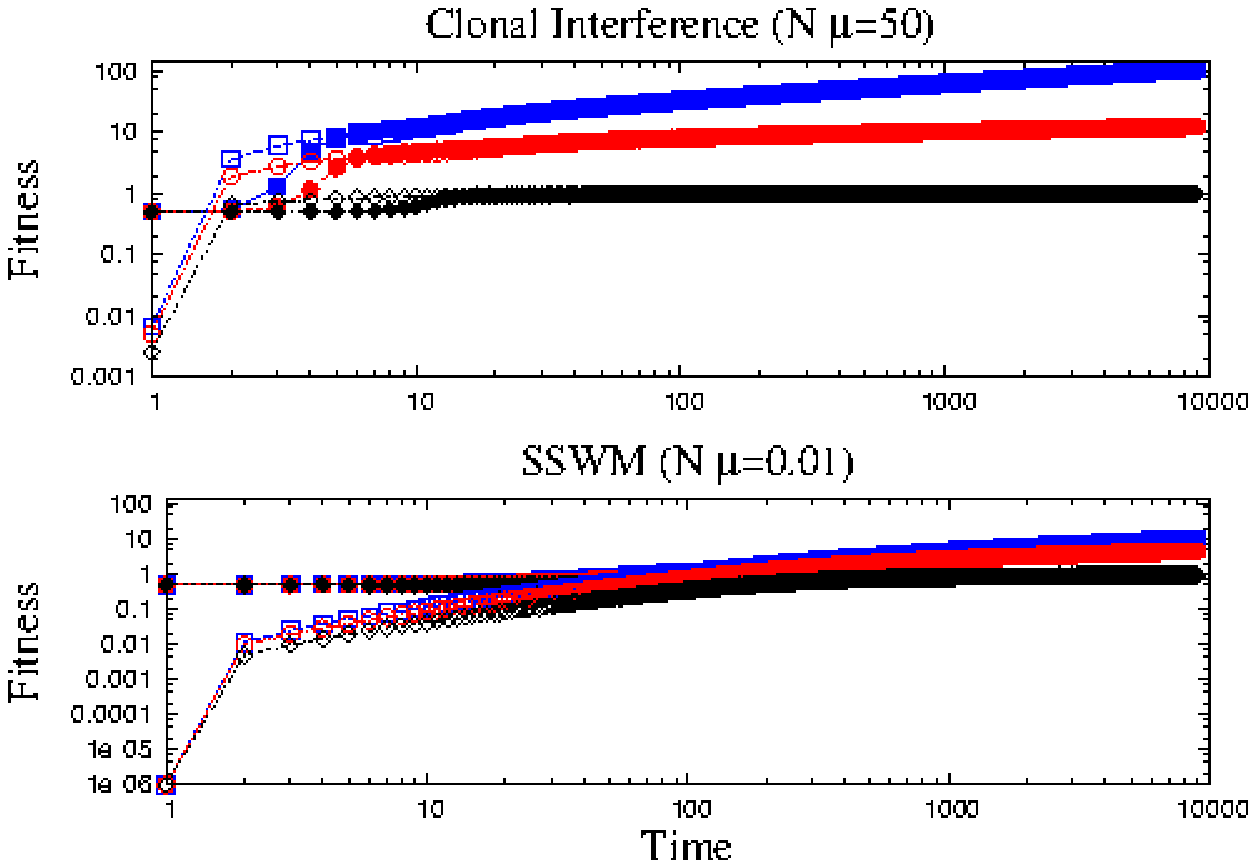}
\caption{\textbf{The plot shows the fitness difference at the first step as a function of the initial 
fitness for different $\kappa$ and two different $N\mu$.} The lines give the theoretical values while the open symbols are 
the simulation output for $N\mu=0.02$ and the closed symbols are those for $N\mu=5$.}  
\label{rf0}
\end{figure}
\begin{figure}[ht]
\renewcommand{\thefigure}{\textbf{S2 Fig.}}
\includegraphics[width=1.0 \linewidth,angle=270]{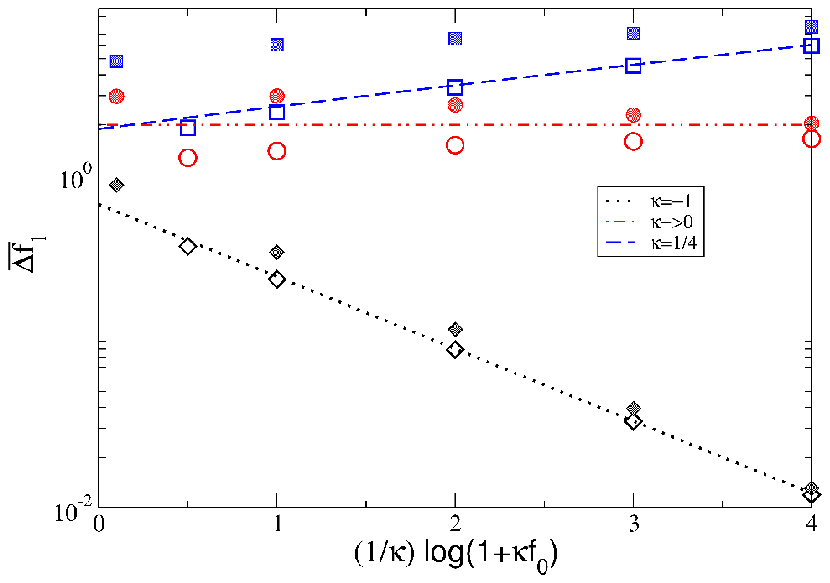}
\caption{\textbf{The figure shows the average fitness of the population for various $\kappa$ in 
both the low and high mutation regimes.} Two different initial conditions $f_0=0$ (open symbols) and $f_0=0.5$ (closed symbols) 
are considered.}   
\label{RA_ic}
\end{figure}
\begin{figure}[ht]
\renewcommand{\thefigure}{\textbf{S3 Fig.}}
\includegraphics[width=1.0 \linewidth,angle=0]{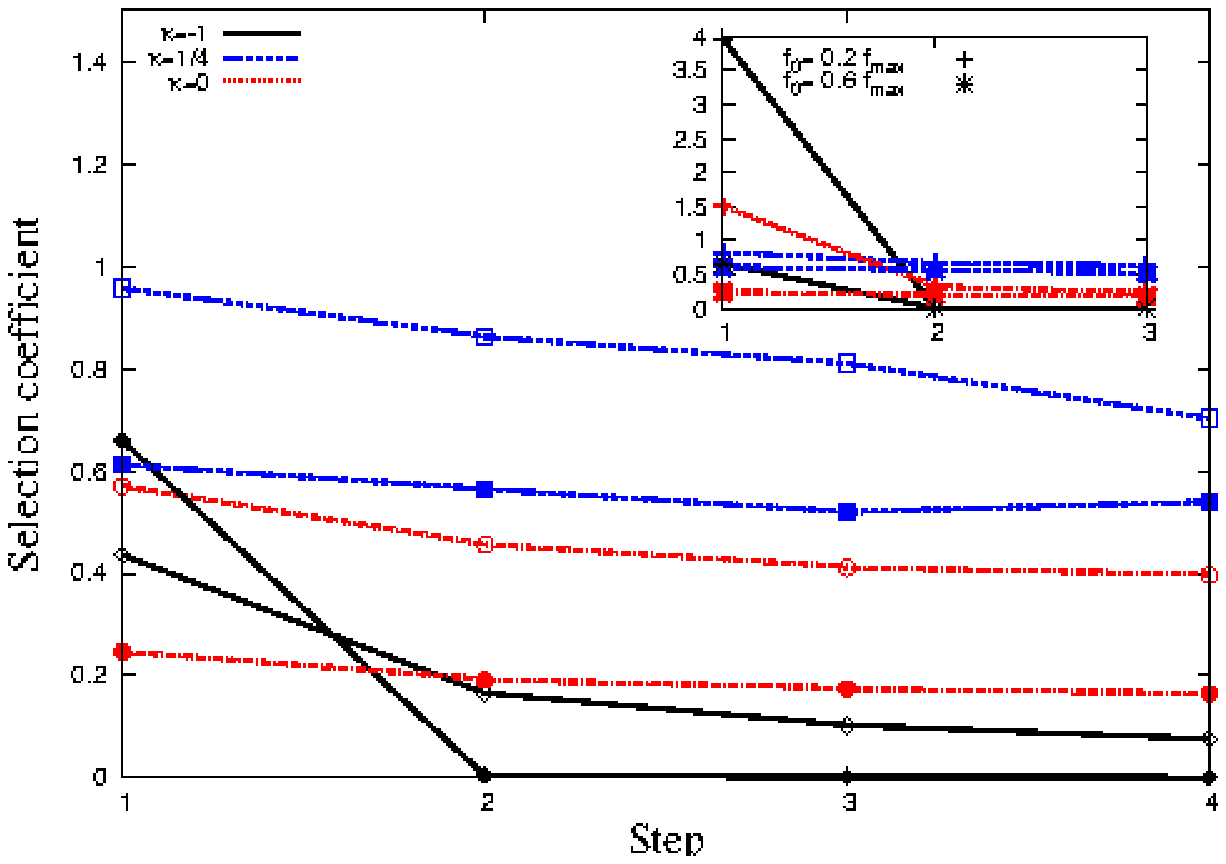}
\caption{\textbf{The main figure shows the selection coefficient as a function of step for 
all three $\kappa$ values}. We considered two different $N \mu$ where open symbols and closed symbols are for $N\mu=0.01$ and 
 $N\mu=50$, respectively. The inset shows the selection coefficient of various steps for two different the initial fitnesses $f_0=0.2 f_{max}$ and $f_0= 0.6 f_{max}$, where $f_{max}$ is calculated using (\ref{fmax}) in the high mutation regime.
}    
\label{fig5}
\end{figure}

\end{document}